\documentclass[sigconf,nonacm]{acmart}

\usepackage{amsmath}
\usepackage{array}
\usepackage{multirow}
\usepackage{textcomp}
\usepackage{xspace}
\usepackage{tcolorbox}
\usepackage{enumitem}

\hypersetup{hidelinks}

\AddToHook{shipout/after}{%
  \ifnum\value{page}=\numexpr\getrefnumber{TotPages}-1\relax
    \balance
  \fi
}

\setcopyright{none}
\acmConference{}{}{}
\acmBooktitle{}
\acmDOI{}
\acmISBN{}
\newcommand{\tool}{\textsf{RAIN}\xspace}

\newcommand{\hide}[1]{}

\begin{document}

\title[RAIN: Agentic Proof Generation for Rocq-based Software Verification]{\tool: Agentic Proof Generation for Rocq-based Software Verification}

\author{Jinyi Zhou}
\affiliation{%
  \department{State Key Laboratory of Novel Software Technology}
  \institution{Nanjing University}
  \city{Nanjing}
  \country{China}
}
\author{Zenan Li}
\affiliation{%
  \institution{ETH Zurich}
  \city{Zurich}
  \country{Switzerland}
}
\author{Yuan Yao}
\affiliation{%
  \department{State Key Laboratory of Novel Software Technology}
  \institution{Nanjing University}
  \city{Nanjing}
  \country{China}
}
\author{Jun Ma}
\affiliation{%
  \department{State Key Laboratory of Novel Software Technology}
  \institution{Nanjing University}
  \city{Nanjing}
  \country{China}
}
\author{Taolue Chen}
\affiliation{%
  \institution{Birkbeck, University of London}
  \city{London}
  \country{United Kingdom}
}
\author{Xiaoxing Ma}
\affiliation{%
  \department{State Key Laboratory of Novel Software Technology}
  \institution{Nanjing University}
  \city{Nanjing}
  \country{China}
}
\renewcommand{\shortauthors}{Zhou et al.}

\begin{abstract}
Interactive theorem provers can provide strong correctness guarantees for software, but constructing the proofs remains a challenging and labor-intensive task. 
%
Inspired by the recent progress of modern coding agents, we present \tool, a lightweight proof-specialized agent for Rocq proof generation. \tool augments an LLM with proof engineering capabilities, including repository exploration, iterative refinement, and checker feedback. 
This design allows the agent to retrieve lemmas autonomously, revise complete proof scripts instead of isolated tactics, and iteratively refine its proof strategy using enriched proof-state information.
Evaluated on a subset from CoqStoq, \tool shows substantial improvements over the existing baselines, 
%
achieving over 90\% prove success rate with at least 24.7\% improvements. We also compare with general-purpose coding agents, the results show that \tool maintains comparable prove success rate but with significant lower cost. 


\end{abstract}


\maketitle

\section{Introduction}


Software bugs remain a central risk in safety- and security-critical systems. Recent infrastructure failures keep this risk concrete: a faulty CrowdStrike update crashed Windows hosts and disrupted airlines and airport operations worldwide in 2024~\cite{milmo2024windows}, while a Cloudflare outage made major online services unreachable or return errors in 2025~\cite{sabin2025cloudflare}. Testing and static analysis are indispensable, but they provide assurance over selected executions, abstractions, or classes of bugs. For systems where stronger guarantees are required, formal verification offers a different standard: the desired behavior is stated mathematically, and the correctness is established by a machine-checkable proof.

Interactive theorem provers (ITPs) have become an essential tool for providing such strong assurance. ITPs such as Rocq (formerly known as Coq), Isabelle/HOL and Lean have supported a plethora of verification projects, including the landmark formal proof of the four-color theorem~\cite{gonthier2008formal}, the verified C compiler CompCert~\cite{leroy2009formal} and the verified microkernel seL4~\cite{klein2009sel4}. 
The obstacle is human proof engineering: developers must understand the verified program, the ITP and the proof techniques needed to connect the implementation with its specification. Indeed, proof developments in large verification projects are often comparable to, or substantially larger than, the implementations they verify \cite{leroy2009formal,klein2009sel4,sanchez2020generating,lu2024proof}. Reducing proof-construction cost is therefore essential to make formal verification practical at scale.

Recently, learning-based systems have shown remarkable progress in automating ITP proof generation. For example, ProverBot9001~\cite{sanchez2020generating} and Tactician~\cite{blaauwbroek2020tactician} use proof corpora to recommend tactics. 
LLM-based systems further combine proof generation with checking, retrieval, repair, decomposition, and search~\cite{first2023baldur,lu2024proof,thompson2024rango,kasibatla2024cobblestone,thakur2023context}. 
However, several limitations remain for repository-level Rocq development. Whole-script generation can preserve a global proof plan, but the first generated script may fail because of invalid references, wrong rewrite directions, focus mistakes, tactic misuse, or other local errors. Rule-based repair can address some surface symptoms, but it gives little help when the proof must choose a different lemma, adjust the case structure, or revise the proof idea itself \cite{lu2024proof}. Tactic-by-tactic systems expose each intermediate proof state, but long proofs face accumulated errors, depth bias, and local choices that may lead to unproductive decompositions \cite{xin2025bfs,he2026stepwise}. Retrieval-augmented systems improve context selection, but fixed or similarity-ranked retrieval can miss decisive project facts \cite{lu2024proof,thompson2024rango,kasibatla2024cobblestone,sun2026reasoning}. 
These limitations point to the need for a workflow that tightly integrates repository exploration, iterative refinement, and checker feedback.


Modern tool-using coding agents directly address this need. 
Given a configured proof environment and a proof-generation task, an agent can inspect definitions and lemmas across the repository, read existing proof scripts, edit the target proof, invoke the proof checker, interpret its feedback, and iteratively refine the proof within the same project context \cite{opencode,anthropic2026claudecode,openai2026codex,yao2022react}. This workflow closely mirrors manual proof engineering, 
and therefore offers a natural way to combine ITP-based verification with LLM-based automation. 


However, simply connecting a general-purpose coding agent to the Rocq checker is insufficient. Efficient proof development depends on rich feedback: the failing sentence, current goals, local hypotheses, relevant project definitions, and errors in later proof branches can all determine the next useful revision. A generic compiler-style error message alone omits much of the proof state needed to select appropriate lemmas, choose rewrite directions, restructure cases, or revise the proof strategy. Consequently, the design focus shifts from isolated tactic generation to a repository-aware, feedback-rich workflow. 

\smallskip
\noindent{\bf Approach.}
%
We develop \tool{}, a lightweight proof-specialized agent for Rocq proof generation. Rather than replacing the LLM with specialized proof-search algorithms, \tool{} treats proof generation as an agentic proof-engineering task. The LLM remains responsible for repository exploration, proof construction, and semantic repair, while the surrounding system 
provides repository-wide workspace, fine-grained proof validation, and structured proof-state information for efficient proof refinement.
This design allows the agent to retrieve lemmas autonomously, revise complete proof scripts instead of isolated tactics, and iteratively refine its proof strategy using enriched proof-state information. The resulting workflow closely mirrors human proof engineering while preserving Rocq as the trusted proof boundary. We emphasize that in this paper, we focus on Rocq proof generation, but this is largely a matter of engineering convenience. The underlying methodology is general and is expected to apply to
other (tactic-based) ITPs with modest adaptation.

\smallskip
\noindent{\bf Evaluations.}
We evaluate \tool on sampled subsets from CoqStoq~\cite{thompson2024rango} and PnVRocqLib~\cite{kasibatla2024cobblestone} benchmarks, and compare it with recent automated Rocq-proof generation methods, PALM~\cite{lu2024proof} and Cobblestone~\cite{kasibatla2024cobblestone}. 
\tool solves 95.8\% theorems in PnVRocqLib and 
90.3\% in CoqStoq, 
improving Cobblestone and PALM 
by 42.3\% and 24.7\%, respectively. 
We further compare with general-purpose coding agents including Claude Code and Codex. On the sampled CoqStoq theorems, 
\tool maintains comparable proof-generation success rate with significant lower proof-obtaining cost, 
reducing the monetary cost by 48.8\% over Claude Code with DeepSeek-v4-flash and 17.3\% over Codex with GPT-5.4. 
On the recent Parcas repository which is published after the LLM release time, \tool keeps the same advantage: it remains over 90\% prove success rate, while significantly reducing the monetary cost compared with general-purpose coding agents. 





\smallskip
\noindent{\bf Contributions.} This paper makes three contributions. 
\begin{itemize}[leftmargin=*]
    \item We formulate Rocq proof generation as an agentic proof-engineering workflow. 
    The LLM remains responsible for proof construction and repair while interacting with a harness specially designed for proof engineering. 
    
    \item We introduce a state-aware feedback mechanism for agentic proof generation. Beyond raw checker messages, it exposes the local proof state through goals and hypotheses, collects multiple errors in a single checking pass, and integrates them into structured feedback needed for effective proof repair.

    \item  We implement the agentic workflow in \tool, a lightweight 
    Rocq agent that augments LLMs 
    with proof-specialized capabilities. 
    Experiments show that \tool{} achieves a substantially higher proof-generation success rate than prior provers. 
    
\end{itemize}  

\noindent{\em Data Availability}.
The code and evaluation scripts are available at \url{https://github.com/RAIN-Prover/RainProver}.

\section{Preliminaries}



Rocq is a widely adopted interactive theorem prover (ITP) based on dependent type theory. A Rocq project typically contains definitions, functions, inductive datatypes, theorem statements, and machine-checkable proofs~\cite{rocqrefman900}. 
A {\em proof} script in Rocq is usually written as a sequence of {tactics}. A {\em tactic} transforms the focused goal into zero, one, or several new goals. During this process, Rocq maintains a {\em global proof state}: the remaining goals and the local proof state available for each goal. The {\em local proof state} contains hypotheses introduced or derived in the current branch. Tactics can also use the global context, which contains imports, definitions, notations, automation hints, and previously proven lemmas. 
If a lemma name is unavailable, a rewrite direction is wrong, a branch remains unfinished, or another proof error occurs, Rocq rejects the script. 

\begingroup\footnotesize
\begin{verbatim}
Theorem t : <statement>.
Proof.
  <tactic>.
  - <branch tactic>.
  - { <branch tactic>. }
Qed.
\end{verbatim}
\endgroup

A skeleton of Rocq source {\em file} is shown above, which is essentially a sequence of sentences. A {\em sentence} is a complete command accepted by the Rocq parser, usually ending at a period. In the example above, the theorem declaration, \texttt{Proof.}, each tactic command, each bullet or brace, and \texttt{Qed.} are all sentences. 
Rocq uses bullets such as \texttt{-}, \texttt{+}, and \texttt{*} to organize proofs with multiple subgoals \cite{rocqrefman900}. By default, a {\em branch} opened by a bullet must be completed before the checker moves to the next sibling branch. 

\section{Motivation and Design Choices}
\label{sec:rain-motivation}

\tool is motivated by the gaps between existing proof-generation techniques and the demands of repository-level proof generation. We identify four key limitations of existing approaches, each of which motivates a corresponding design choice in \tool. 

\begin{figure*}[t]
\centering
\begin{minipage}[t]{0.48\textwidth}
\centering
\small\textbf{(a) Missing local proof state}\par\vspace{0.25em}
\includegraphics[width=\linewidth]{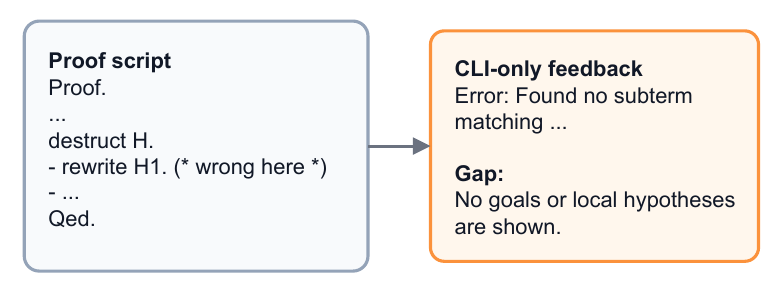}
\end{minipage}\hfill
\begin{minipage}[t]{0.48\textwidth}
\centering
\small\textbf{(b) First-error-only checking}\par\vspace{0.25em}
\includegraphics[width=\linewidth]{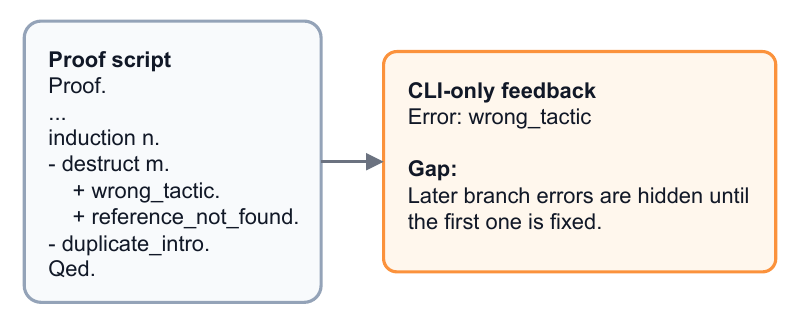}
\end{minipage}
\caption{Two feedback gaps of off-the-shelf general-purpose coding agents. (a) Command-line errors from Rocq do not include the local proof state. In the example, \texttt{rewrite H1.} triggers an error stating that no matching subterm was found. Without the local goals and hypotheses, the LLM must infer whether the rewrite direction, chosen hypothesis, or preceding proof state is wrong. (b) First-error-only checking isolates individual errors without collectively considering later branch failures. In the example, an error in the first branch can hide later-branch errors that may be local mistakes in different branches or evidence of an earlier proof-decomposition problem, forcing the agent to fix the earlier error before it can see the later errors.}
\label{fig:agent-feedback-gaps}
\Description{Two Rocq feedback examples. The first shows a rewrite error without the local goals and hypotheses. The second shows how an error in an earlier branch prevents command-line checking from reporting errors in later branches.}
\end{figure*}

\smallskip
\noindent{\bf Limitation 1: prescriptive lemma retrieval}.
LLM-based proof-generation systems usually receive lemmas using a fixed retrieval strategy, which may omit the key lemma needed by a proof. For example, PALM~\cite{lu2024proof} uses CoqHammer's \texttt{predict} and BM25 reranking; Rango~\cite{thompson2024rango}, ranks available theorem/lemma statements with TF-IDF over identifiers in the current focused goal; Cobblestone~\cite{kasibatla2024cobblestone} supplies the LLM with lemmas that precede the target theorem in the current file in its reported non-oracle setting. 
Such prescriptive retrieval can be brittle: they may rank a generic but decisive lemma outside the selected top-$k$ lemmas, and the file context can miss lemmas defined in other files. 
Once the decisive lemma is excluded from the supplied context, the LLM has little chance to see it during generation, and must either attempt a much harder proof or re-derive a lemma already present in the library.

\textbf{\em Design choice 1: autonomous lemma retrieval}. Inspired by the recent progress in autonomous agents,
\tool lets the LLM itself retrieve lemmas through repository-level tools. After reading the target theorem, the LLM can search and read project files according to its own proof ideas. In particular, after seeing error feedback, the LLM may realize that the current proof lacks a needed lemma, and search the repository to find the missing lemma before revising the proof.

\smallskip
\noindent{\bf Limitation 2: surface proof repair}.
Prior work observes that an LLM-generated proof script may contain a plausible high-level proof idea while still failing because of local mistakes, such as wrong rewrite directions, bullet-level errors, or a bad choice of induction variable. PALM~\cite{lu2024proof} therefore proposes to repair such failures with 
reference replacement, renaming, bullet transformation, premise augmentation, and CoqHammer-based backtracking.
These repairs are useful but limited. Many failures are semantic-level rather than syntactic-level errors. For example, an earlier induction choice may be too strong, or the proof method itself may be wrong. Such cases require revising the proof idea, which is beyond the reach of existing repairs on identifiers, bullets, or tactics. 

\textbf{\em Design choice 2: semantic proof repair.}
For proof repair, our key insight is that error messages and Rocq's internal states can effectively guide the LLM in repairing failed proofs. \tool therefore keeps repair inside the LLM-driven proof loop: after seeing checker feedback and internal states (see design choice 3), the LLM can reason about the failure semantically and revise the proof idea itself by, e.g., changing an overly strong induction choice or switching to a different proof method.

\smallskip
\noindent{\bf Limitation 3: insufficient feedback}. General-purpose coding agents such as Claude Code and Codex are also strong proof-generation baselines. 
However, when directly using them with the default 
command-line Rocq check, the feedback available to the agent is usually insufficient to efficiently generate the proofs. 
As illustrated in 
Figure~\ref{fig:agent-feedback-gaps}, we summarize two types of missing feedback: (1) {\em missing local proof state}, where local goals and hypotheses are not exposed in command-line errors, making LLM struggle with inferring whether the rewrite direction is wrong, the wrong hypothesis was chosen, or the earlier tactic produced a wrong proof state; (2) {\em first-error-only checking}, where the agent checks and fixes the first error before proceeding to later errors, while multiple errors may be correlated and considering them together can better steer the proof repair. 

\textbf{\em Design choice 3: feedback augmentation.}
We augment the feedback in two ways: (1) {\em local-state collection}, and (2) {\em multi-error collection}.
Specifically, \tool first records the local proof state just before the failing sentence and returns it in the error messages for LLMs (Cf. Section~\ref{sec:rain-environment-capture}). 
It then uses temporary Rocq admitting to skip the earlier failing branch, allowing errors in later branches to be returned in the same feedback packet (Cf. Section~\ref{sec:error}). 

\smallskip
\noindent{\bf Limitation 4: deficiency in long-range proofs}.
Finally, several recent systems generate proofs tactic by tactic. BFS-Prover~\cite{xin2025bfs} and the seL4 neuro-symbolic prover~\cite{he2026stepwise} expand proof states step by step and prune invalid prefixes using checker feedback. This tactic-by-tactic proving gives the LLM immediate proof-state feedback, but it also faces difficulties on long proofs: errors accumulate along the trajectory, search scores can bias against deep paths, and locally valid tactics may lead the proof into an unproductive decomposition. 
For example, the seL4 neuro-symbolic prover identifies a significant performance drop in long-proof theorems: the proof success rate for 393 theorems with ground-truth proofs exceeding 10 lines is around 20\%, in comparison with the average 77\% success rate.

\textbf{\em Design choice 4: whole-script editing.}
\tool adopts a whole-script editing strategy, aiming to produce a global proof plan encoded by complete Rocq proof scripts. Specifically, the LLM can freely modify the target proof, ranging from revising the overarching proof structure or rewriting a large part of the proof plan to making a small local repair, without being confined to a tactic-by-tactic framework.

\section{Approach}
\label{sec:rain}

\begin{figure*}[t]
\centering
\includegraphics[width=0.9\textwidth]{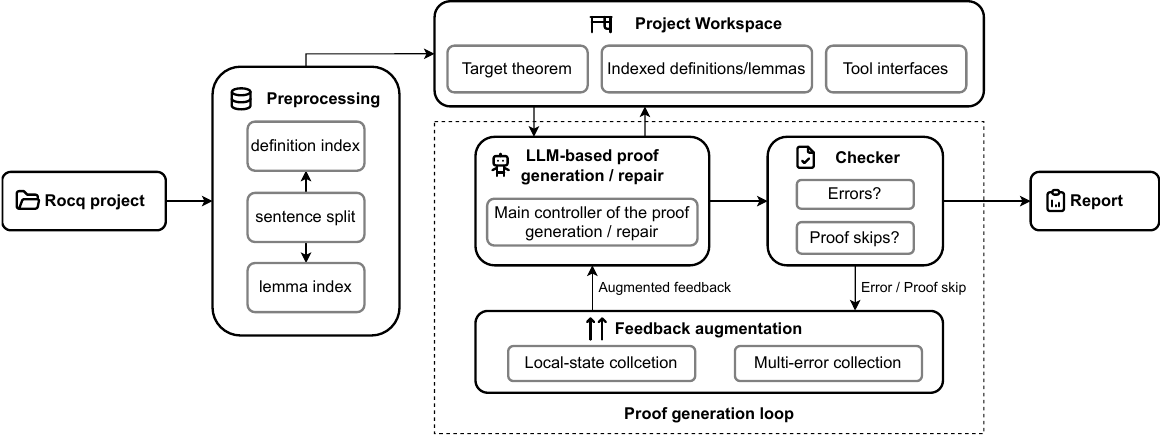}
\caption{Workflow of \tool}
\label{fig:proofagent-workflow}
\Description{The RAIN workflow prepares a Rocq project workspace, then repeats repository search, source reading, proof-script editing, and checker execution. Failed checks return local proof states and errors from multiple branches to guide the next revision.}
\end{figure*}

This section presents \tool, a lightweight specialized agent for Rocq proof generation. \tool keeps the LLM responsible for proof construction and repair, and surrounds it with a repository-aware workspace, 
fine-grained checker execution, and augmented feedback for iterative refinement. The method is intentionally organized around the proof-engineering loop that a human developer would follow: inspect project context, edit the proof script, run Rocq, read the failure state, and revise the proof. Figure~\ref{fig:proofagent-workflow} summarizes this workflow, which consists of two major stages.

\begin{itemize}[leftmargin=*]
    \item \textbf{Preprocessing.} \tool first prepares the project workspace used by the agent. It validates that the target project can be checked before proof generation, splits Rocq files into parser-level sentences, extracts definitions related to the target theorems, and indexes lemma and theorem statements in the project. The resulting workspace gives the LLM a compact initial context and a set of scoped repository operations for later search and reading.
    \item \textbf{Proof generation loop.} \tool organizes proof generation as a `search-read-edit-feedback' loop. The LLM searches for relevant names, lemmas, definitions, or proof patterns, reads selected statements or source ranges, edits the complete target proof region, and invokes the Rocq checker. When Rocq reports a failure, \tool augments the checker output with local-state collection and multi-error collection before returning it to the LLM. The LLM then uses the feedback to decide whether to gather more repository context, make a local repair, or rewrite a larger part of the proof. In this loop, the LLM controls the proof strategy, while each candidate proof is checked by Rocq and each failed check provides feedback for the next revision.
\end{itemize}

This structure instantiates the design choices in Section~\ref{sec:rain-motivation}. The project workspace supports autonomous lemma retrieval; whole-script editing lets the LLM revise a complete proof plan; checker execution supports semantic proof repair; and the two feedback mechanisms make checker output more useful for the next revision. The rest of this section describes the preprocessing workspace, local-state collection, and multi-error collection. Because the evaluation needs a precise success criterion for agent-generated proofs, we also describe the validation procedure in Section~\ref{sec:rq2-agents}.

\subsection{Preprocessing}

Preprocessing turns a raw Rocq project into the workspace that \tool exposes to the LLM. The first requirement is reliable sentence splitting. Rocq source files are processed as parser-level sentences, but these sentence boundaries cannot be recovered safely from periods, bullets, or braces alone. Such characters may appear in comments, strings, notation, or compact proof scripts, and a wrong boundary would make later local-state collection or multi-error collection unsafe. \tool therefore delegates sentence splitting to an external parser built from the document-management pipeline used by VsRocq~\cite{vsrocq_placeholder}. The splitter returns ordered sentences with stable identifiers, source spans, source text, and coarse sentence categories. If sentence splitting times out, exits with an error, returns invalid JSON, or produces an empty sentence list, \tool disables the higher-level features that depend on sentence boundaries.

The splitting result helps \tool classify definitions, lemmas, and theorems. More importantly, it lets \tool accurately recover their statements for later proof-generation steps. The same sentence boundaries are also used by local-state collection and multi-error collection to place temporary queries and identify the proof branch associated with an error.

\tool then builds two lightweight indexes. The definition index records the name, source span, and file location of definitions that are relevant to the target theorem; definitions appearing in the target statement are included in the initial knowledge block so the LLM starts from the local vocabulary of the theorem. The statement index records lemma and theorem declarations by file, including each declaration's name, proposition, and source location. This index is designed for proof engineering rather than for one-shot retrieval: the LLM can browse the statements in a selected file, search names or source text across the project, and then read the exact source range around a promising definition or lemma.

After preprocessing, \tool produces a workspace that the LLM can operate on directly in the proof-generation loop. Through this workspace, the LLM can search statement names and source text across the repository, browse the declarations recorded for a file, read the source around a selected definition or lemma, replace the target proof region, run the Rocq checker, and receive augmented feedback from failed checks.

\subsection{Local-State Collection}
\label{sec:rain-environment-capture}

Local-state collection targets the missing local proof state described in Section~\ref{sec:rain-motivation}. Rocq command-line errors often identify the failing sentence but do not show the goals and hypotheses that made the sentence fail. For an LLM editing a complete proof script, this missing state changes the repair problem. The LLM may have to guess whether the chosen hypothesis has the wrong shape, whether the rewrite direction is reversed, whether a previous destruct produced a different branch than expected, or whether the current tactic is aimed at the wrong subterm.

Proof-skipping behavior creates a related feedback need. A normal Rocq check does not report an error at proof-skipping sentences such as \texttt{Admitted.} or \texttt{Abort.}, but these sentences still mark a place where the LLM stopped without completing the proof. \tool therefore also treats the skip location as a place where local proof state should be recovered, so the next revision can see the goal that was left unsolved.

For both checker errors and proof-skipping sentences, the key idea is to recover the proof state immediately before the relevant Rocq sentence. That is the state Rocq would use to execute the failing sentence or the skip sentence. All previous sentences have already shaped the proof state, and the failing or skip sentence has not yet changed it. \tool augments the feedback with the goals and hypotheses at that point, so the LLM can repair the proof using the actual state that led to the failure or skip.

Operationally, local-state collection reruns sentence splitting on the file that contains the checker error or detected skip. \tool uses the splitting result to locate the Rocq sentence at that position, creates a temporary edit that inserts a proof-state query immediately before that sentence, and reruns the configured checker. The captured goals and hypotheses are returned to the LLM together with the original checker error or skip report. The original source remains the proof attempt being repaired; the temporary query is used only to obtain feedback. If Rocq cannot reach the query or the query fails to produce a state, \tool still returns the original checker error or skip report.

This feedback gives the LLM concrete proof-state evidence for the next whole-script edit. A failed rewrite can be compared with the current goal term; a failed application can be checked against the hypotheses in scope; and an unfinished branch can be diagnosed from the remaining goal. The LLM can then repair the proof semantically, for example by choosing a different hypothesis, changing the rewrite direction, unfolding a missing definition, or changing the local proof structure. If the captured state shows that an earlier proof decomposition led to an unexpected goal, the LLM can revise that earlier decomposition instead of repeatedly editing the failing sentence.

\subsection{Multi-Error Collection}
\label{sec:error}

Multi-error collection targets the first-error-only checking behavior described in Section~\ref{sec:rain-motivation}. Rocq stops at the first error, which is the right behavior for proof checking but a narrow feedback interface for proof repair. In a script with several branches, the first branch may fail because a rewrite is aimed at the wrong goal, while later branches already contain an undefined lemma, an invalid introduction, or a tactic written for a different branch shape. With only ordinary checker output, the LLM has to fix the first error before Rocq can reveal the next one. As a result, one proof attempt may require several edit-check cycles just to expose errors in different bullet branches that were already present from the beginning.

The key idea is to use the first error as a reliable anchor and then ask Rocq for more information about the remaining proof text. After recording the first checker error, \tool temporarily skips the branch containing that error with Rocq's admitting mechanism. Rocq can then continue checking later proof text and may expose another error from a different branch. Each collected error can also be augmented by local-state collection, so the LLM can see the corresponding goals and hypotheses for several errors. \tool repeats this process up to its collection budget and returns the collected errors together to the LLM. The skipping described above is temporary and is used only to obtain feedback; the LLM still has to produce a proof script that Rocq accepts without skipped branches.

During this collection process, Rocq can also report misleading messages. An incomplete skipped branch may cause Rocq to report an incomplete proof at a proof-closing sentence, and a ``No such goal'' message can indicate that later proof text is no longer synchronized with the current proof state. Errors after such messages may be unreliable, so the collection stops after them. \tool filters these misleading errors instead of returning them to the LLM.

Collecting multiple branch errors is useful in two ways. First, errors in different bullet branches can be repaired in one revision. The LLM can fix a wrong rewrite direction in one branch, replace a missing lemma in another branch, and adjust an invalid introduction elsewhere without waiting for separate checker runs to reveal each problem. Second, errors from different branches are sometimes related. If several branches fail with similar goal shapes or incompatible local contexts, the LLM can infer that the earlier proof decomposition may be unhelpful. It can then change the case split, strengthen the induction hypothesis, move a common simplification before branching, or replace branch-by-branch reasoning with a more uniform proof plan.

The implementation again relies on parser-level sentences. To choose the skipped region, \tool simulates Rocq's bullet and brace structure up to the failing sentence, which identifies the nesting level of the branch containing the error. It then continues through later sentences until the proof leaves that nesting level; the previous sentence is treated as the end of the failing branch. For each temporary checking run, \tool comments out the text from the error location to the end of that branch and inserts \texttt{all:\ admit.}, allowing Rocq to continue checking and report later errors. This resembles Cobblestone's fail-safe execution, which also uses temporary admitting to continue execution after errors and localize valid subproofs~\cite{kasibatla2024cobblestone}, while \tool uses the idea to enrich feedback for the next LLM revision.

\subsection{Fine-Grained Validation}
\label{sec:rain-validation}

After any proof-generating agent stops, its generated repository still needs an explicit acceptance check. The agent's final message is not a proof-success signal because it does not show that the target file contains a complete checked proof. The generated repository must first pass Rocq checking; this is the condition that the proposed proof is accepted by the proof assistant. Checker success alone leaves three additional issues for agent outputs: an agent could change the target statement and prove a different theorem, proof-skipping commands such as \texttt{Admitted.} and \texttt{Abort.} do not produce checker errors, and an agent could introduce new axiom-like declarations elsewhere in the repository. We therefore validate agent-generated proofs as follows.
\begin{itemize}[leftmargin=*]
    \item The verifier first runs the project checker on the generated repository. The generated script must be accepted by Rocq in the same project context in which the target theorem is evaluated. If the checker fails, the generated proof is rejected regardless of the agent's final message.
    \item The verifier also checks that the target theorem has not been weakened or replaced. It records the target statement before proof generation and compares it with the statement after the agent stops. A generated proof is rejected if the theorem statement changes, even when the modified project can be checked. This prevents a proof attempt from succeeding by proving a different or easier theorem.
    \item Finally, the verifier rejects shortcuts that bypass the intended proof. It scans the target file outside comments and strings for proof-skipping commands, and it checks that the generated repository does not introduce new axiom-like declarations relative to the prepared input. 
\end{itemize}

These checks rule out common ways for an agent to make the project pass without completing the intended proof. A generated proof is accepted only when all verifier checks pass.

\section{Evaluation}

We evaluate \tool, focusing on the following four research questions (RQs):

\begin{description}
\item[RQ1:]
How does \tool perform against existing automated provers? 
\item[RQ2:]
How does \tool perform against general-purpose coding agents?
\item[RQ3:]
How do \tool's proof-specific feedback modules contribute to its performance?
\item[RQ4:]
How well does \tool generalize to 
other Rocq projects?
\end{description}

\newcommand{\rqresponse}[2]{%
\par\smallskip
\noindent\fcolorbox{black!35}{black!5}{%
\begin{minipage}{\dimexpr\linewidth-2\fboxsep-2\fboxrule\relax}
\small\textbf{Response to #1.} #2
\end{minipage}}%
\par\smallskip }

\subsection{Experimental Setup}
\label{sec:evaluation-setup}

\noindent{\bf Benchmarks}.
Our primary benchmark consists of 100 theorems sampled uniformly at random from the public CoqStoq test split introduced by Rango~\cite{thompson2024rango}. This split contains real open-source Rocq projects configured for proof-synthesis evaluation. The targets retain project dependencies, notation, and proof conventions, matching the conditions of realistic Rocq proof development.

For the recent-project generalization experiment in RQ4, we additionally use Parcas~\cite{moine2026parcas}, a recent Rocq development, and randomly sample 100 target proofs from it. This repository was first uploaded on May 29, 2026, after the April 24, 2026 release of the DeepSeek model used in that experiment. 

When the evaluated system is an agent, for each target theorem, we replace the proof body with \texttt{Admitted} while preserving the theorem statement, imports, surrounding definitions, and project build configuration. The agent is then asked to complete the proof in the target file, without editing theorem statements or using proof-skipping constructs such as \texttt{Admitted}, \texttt{Abort}, and \texttt{Axiom}.

\smallskip
\noindent{\bf Models}.
We consider three LLMs in our experiments, namely, DeepSeek-v4-flash~\cite{deepseek2026pricing}, Qwen3.5-Flash~\cite{qwen3535ba3b}, and GPT-5.4~\cite{openai2026models}. We use DeepSeek-v4-flash by default.
DeepSeek-v4-flash runs with its reasoning effort set to max. 
Qwen3.5-Flash runs under its thinking mode. GPT-5.4 runs use reasoning effort xhigh.

\smallskip
\noindent{\bf Baselines}.
We mainly compare \tool with Cobblestone~\cite{kasibatla2024cobblestone} and PALM~\cite{lu2024proof}. 
The comparisons use the same backend model whenever the baseline makes LLM calls. We also include the reported results of ReCent-Prover~\cite{sun2026reasoning} for comparison.\footnote{Its implementation is not publicly available.} We do not reproduce earlier systems such as ProverBot9001~\cite{sanchez2020generating} and Tactician~\cite{blaauwbroek2020tactician} because they were already included in the evaluations of Cobblestone or PALM and were significantly outperformed by these two recent baselines on the corresponding benchmarks.
Additionally, we also compare with several off-the-shell coding agents including Claude Code, Codex, and OpenCode.

\smallskip
\noindent{\bf Evaluation Metrics}.
We first adopt the {\em prove success rate} as the key effectiveness metric. It measures the pass rate of the generated proofs. By default, We consider the success rate at pass@$1$, and further report pass@$k$ results for stability analysis. 

For efficiency, we evaluate both monetary cost and wall-clock time. Monetary cost is computed from the billable input and output tokens, including cache reads when reported by the provider. For both DeepSeek-v4-flash and GPT-5.4, we use the official pricing model, which distinguishes among cache-hit input tokens, cache-miss input tokens, and output tokens. 
For Qwen3.5-Flash, the monetary cost columns are replaced by total-token usage in millions of tokens, as the API does not reliably return cache-usage fields.
The intuition behind the above monetary cost measurement is that, 
for agentic workflows, they typically append new messages and tool-use results to the conversation history, and the existing prompt prefix remains unchanged across turns. 

\smallskip
\noindent{\bf Configurations}.
The \tool system prompt is adapted from OpenCode's public coding-agent prompt~\cite{opencode}. Each proof attempt has a 30-minute time limit. 
During multi-error collection, \tool is configured to collect up to two additional later checker errors after the first failing error, and each checker invocation uses a 60-second timeout. \tool can read only files inside the target project for most experiments, 
and is allowed to inspect dependency source files in RQ4. 

All experiments were run on a Linux server running Ubuntu 22.04.5 LTS, with an AMD Ryzen Threadripper PRO 5975WX CPU (32 cores and 64 hardware threads), one NVIDIA GeForce RTX 3090 GPU, and 128GB RAM.

\subsection{RQ1: Comparison with Automated Provers}

\noindent{\bf Comparison with Cobblestone}.
For fair comparison, we use the PnVRocqLib benchmark in this experiment because it is used in Cobblestone's own evaluation and supports our Coq version 8.18. 
We use the same 100 targets selected in Cobblestone's evaluation.
We run Cobblestone's strongest fully automatic non-oracle setting. 
To favor Cobblestone and show its capability limit, we raise its context limit to fit the model context window, give it three times \tool{}'s time limit (i.e., 90 minutes per theorem), and provide a large expansion-node budget. 
Both systems encounter some targets that fail during their own preprocessing, due to benchmark/source drift or parser limitations in the versions used for evaluation. We therefore report each system on the targets that pass its own pre-search initialization, and also report the common valid-target set (all Cobblestone-valid targets pass \tool{}'s initialization).

\begin{table}[t]
\centering
\footnotesize
\caption{Comparison against Cobblestone on PnVRocqLib.}
\label{tab:rq1-cobblestone}
\setlength{\tabcolsep}{2pt}
\begin{tabular*}{\columnwidth}{@{\extracolsep{\fill}}lccccc@{}}
\toprule
  &  \multirow{2}{*}{Time limit} & \multicolumn{2}{c}{Separate set} & \multicolumn{2}{c}{Shared set} \\
\cmidrule(lr){3-4}\cmidrule(l){5-6}
 & & \shortstack{Success\\rate} & \shortstack{Absolute\\gain} & \shortstack{Success\\rate} & \shortstack{Absolute\\gain} \\
\midrule
\tool & 30m & 93/97{\scriptsize\,(95.9\%)} & +42.4\% & 68/71{\scriptsize\,(95.8\%)} & +42.3\% \\
Cobblestone & 90m & 38/71{\scriptsize\,(53.5\%)} & -- & 38/71{\scriptsize\,(53.5\%)} & -- \\
\bottomrule
\end{tabular*}
\end{table}

Table~\ref{tab:rq1-cobblestone} shows the proof success rate results; absolute gain denotes \tool{}'s success-rate improvement over the baseline. 
Cobblestone successfully generates proofs for 53.5\% of the benchmark targets,\footnote{The reported success rate is around 44\% in the Cobblestone paper.}
while \tool solves over 95\% problems in both the separate set and the shared set, achieving at least 42.3\% success-rate gain.

\smallskip
\noindent{\bf Comparison with PALM}.
PALM's original evaluation uses CoqGym with Coq 8.10, 8.11 and 8.12. These versions are below the minimum Coq version supported by \tool. We therefore port PALM to Coq 8.18 and evaluate it on the same 100 randomly sampled CoqStoq tasks. To favor PALM, we start from its original pass@20 setting with 30 selected lemmas and then give unsolved tasks a second pass@20 round with the selected-lemma budget increased to 100.\footnote{We increase the budget because PALM's own failure analysis identifies missing premises as the most frequent failure category, accounting for 58\% of sampled failures.} 7 theorems cannot be evaluated reliably in the PALM environment, so PALM is evaluated on 93 tasks. For \tool, we report one DeepSeek-v4-flash run on the same samples (i.e., pass@1).

\begin{table}[t]
\centering
\footnotesize
\caption{Comparison against PALM on CoqStoq.}
\label{tab:rq1-palm}
\setlength{\tabcolsep}{2pt}
\begin{tabular*}{\columnwidth}{@{\extracolsep{\fill}}lccccc@{}}
\toprule
 & \multirow{2}{*}{Trials} & \multicolumn{2}{c}{Separate set} & \multicolumn{2}{c}{Shared set} \\
\cmidrule(lr){3-4}\cmidrule(l){5-6}
 & & \shortstack{Success\\rate} & \shortstack{Absolute\\gain} & \shortstack{Success\\rate} & \shortstack{Absolute\\gain} \\
\midrule
\tool & 1 & 91/100{\scriptsize\,(91.0\%)} & +25.4\% & 84/93{\scriptsize\,(90.3\%)} & +24.7\% \\
PALM & 40 & 61/93{\scriptsize\,(65.6\%)} & -- & 61/93{\scriptsize\,(65.6\%)} & -- \\
\bottomrule
\end{tabular*}
\end{table}

Table~\ref{tab:rq1-palm} shows the success rate results. 
On the shared set,
PALM solves 61/93 tasks after 40 trials, while \tool solves 84/93 tasks with one trial, achieving a 24.7\% absolute success-rate gain. On the separate set, \tool solves 91/100 tasks, still achieving over 90\% success rate. 



\smallskip
\noindent{\bf Comparison with ReCent-Prover}.
For ReCent-Prover, we directly compare with the results reported in its paper. On a random sample of 200 CoqStoq theorems, ReCent-Prover proves 138/200 (69.0\%) with o4-mini and reports 66.7K average online tokens per theorem. On our independently sampled 100 CoqStoq tasks, one DeepSeek-v4-flash run of \tool{} solves 91/100 tasks, with mean and median token counts of 3.05M and 403K per theorem. Thus, \tool{} shows a higher success rate, while ReCent-Prover reports substantially lower online token consumption. These token numbers are only qualitatively comparable because ReCent-Prover's reported per-theorem number excludes the LLM token cost of constructing its retrieval database, and the two systems use different backend models and token-cache distributions.

\begin{tcolorbox}[
  colback=gray!5,
  colframe=black!60,
  title=Response to RQ1
]
\tool significantly improves the proving capability over the existing automated provers, achieving over 90\% prove success rate in multiple benchmarks. 
\end{tcolorbox}



\subsection{RQ2: Comparison with  General-purpose Coding Agents}
\label{sec:rq2-agents}

We compare \tool with general-purpose coding agents under the same instruction. For DeepSeek-v4-flash and Qwen3.5-Flash backends, we combine them with Claude Code~\cite{anthropic2026claudecode} and OpenCode~\cite{opencode}. For the GPT-5.4~\cite{openai2026models} backend, we directly compare with Codex~\cite{openai2026codex}.  \hide{The system and user prompts follow the construction summarized in Section~\ref{sec:evaluation-setup}.}

\smallskip
\noindent{\bf Performance and Stability Analysis}. 
Since agents are relatively unstable across multiple runs,
to reduce variance, we run each theorem 10 times for each agent and report both theorem-level pass@10 and trial-level success rates over all attempts.


Table~\ref{tab:rq2-overall-success} reports the performance of \tool and the baseline coding agents over the sampled CoqStoq benchmark. It can be observed that general-purpose coding agents already perform very well on the benchmark theorems, and agents with strong back-end LLMs perform relatively stable across multiple trials. \tool obtains higher pass@10 and trial-level success than Claude Code and OpenCode with DeepSeek-v4-flash and Qwen3.5-Flash. With GPT-5.4, Codex has a slightly higher success rate. 

\begin{table}[t]
\centering
\footnotesize
\caption{Comparison against general-purpose coding agents on CoqStoq.} 
\label{tab:rq2-overall-success}
\begin{tabular}{@{}llcc@{}}
\toprule
Backend & System & Pass@10 & Trial Success\tabularnewline
\midrule
\multirow{3}{*}{DeepSeek-v4-flash} & \tool & 94.0\% & 89.8\%\tabularnewline
 & Claude Code & 93.0\% & 85.7\%\tabularnewline
 & OpenCode & 91.0\% & 86.3\%\tabularnewline
\midrule
\multirow{3}{*}{Qwen3.5-Flash} & \tool & 55.0\% & 36.7\%\tabularnewline
 & Claude Code & 51.0\% & 24.2\%\tabularnewline
 & OpenCode & 48.0\% & 26.4\%\tabularnewline
\midrule
\multirow{2}{*}{GPT-5.4} & \tool & 95.0\% & 92.7\%\tabularnewline
 & Codex & 96.0\% & 94.7\%\tabularnewline
\bottomrule
\end{tabular}
\end{table}

\begin{table*}[t]
\centering
\footnotesize
\caption{Cost comparison with general-purpose coding agents.} 
\label{tab:rq2-shared-solved}
\begin{tabular}{@{}clc|ccc|ccc@{}}
\toprule
{Backend} & Baseline & Theorems & \multicolumn{3}{c|}{Monetary cost} & \multicolumn{3}{c}{Time cost} \\
\cmidrule(lr){4-6}\cmidrule(l){7-9}
 & & & \tool & Baseline & Reduction & \tool & Baseline & Reduction \\
\midrule
\multirow{2}{*}{DeepSeek-v4-flash} & Claude Code & 93 & \$0.0142 & \$0.0277 & 48.8\% & 338.7s & 448.1s & 24.4\%\\
 & OpenCode & 91 & \$0.0130 & \$0.0196 & 33.5\% & 313.2s & 353.0s & 11.3\%\\
\midrule
\multirow{2}{*}{Qwen3.5-Flash} & Claude Code & 41 & 2.43 M tok & 5.06 M tok & 52.1\% & 453.0s & 784.2s & 42.2\%\\
 & OpenCode & 44 & 2.63 M tok & 5.11 M tok & 48.5\% & 476.3s & 702.3s & 32.2\%\\
\midrule
GPT-5.4 & Codex & 95 & \$0.515 & \$0.622 & 17.3\% & 214.0s & 210.1s & -1.9\%\\
\bottomrule
\end{tabular}\\
\smallskip
{\scriptsize For Qwen3.5-Flash, the API does not reliably return
cache-usage fields, so monetary cost columns are replaced by total-token usage; M tok
denotes million tokens.}
\end{table*}

\smallskip
\noindent{\bf Cost Comparison}.
In addition to the proof success rate, we also measure monetary cost and time cost on the theorem sets where both systems can obtain an accepted-proof at least once. Shared-solved subsets are defined pairwise. For \tool and one baseline agent, we select theorems where both systems succeed at least once out of 10 trials. 
For each selected theorem, we compute each agent's average monetary cost or token usage, and wall-clock time over its 10 attempts, and then average over all the theorems.

Table~\ref{tab:rq2-shared-solved} reports the results, with cost reductions computed relative to the baseline agent. It can be observed that for the solved theorems, \tool incurs significantly lower monetary cost or token usage compared to the general-purpose agents. This is also consistent with the intuition, since \tool is specially designed for theorem proving. 
For wall-clock time, \tool still significantly shortens the running time, with an exception from Codex. The reasons are two-fold. First, \tool{} spends extra time for preprocessing the Rocq project, such as constructing the indexes for definitions and lemmas. Although this step can be done before running the proof loop, we rebuild them for every trial to avoid cross-trial information leakage.
Second, GPT-5.4 often generates an initial proof script that is already close to the valid one, especially for easier theorems. This makes the preprocessing time more dominant in the comparison.

\begin{tcolorbox}[
  colback=gray!5,
  colframe=black!60,
  title=Response to RQ2
]
General-purpose coding agents already perform very well on theorem proving, and \tool achieves close success rates. For weaker LLM backends, \tool is even better. Additionally, due to its specialized design for theorem proving, \tool significantly reduces the monetary cost or token usage of invoking LLMs across all the combinations. 
\end{tcolorbox}


\subsection{RQ3: Ablation Study}
We conduct an ablation study by removing the two feedback modules: local-state collection and multi-error collection. The ablation asks whether these modules improve proof-generation success and whether they reduce wasted repair costs.
We run each ablation variant using DeepSeek. The other setups follow RQ2. For example, the cost comparison is based on the shared-solved subset. 
Table~\ref{tab:ablation-main} shows the results, where `w/o local-state' means removing the local-state collection module, `w/o multi-error' means removing the multi-error collection module, and `w/o both' means removing both. 

\smallskip
\noindent{\bf Local-state collection}. Removing local-state collection reduces pass@10 from 95.0\% to 94.0\% and lowers trial-level success from 90.9\% to 88.0\%. As the cost comparison, 
\tool has 13.0\% lower monetary cost and 26.8\% lower time cost than the no-local-state variant. When multi-error collection is absent, keeping local-state collection still reduces monetary cost by 21.4\% and wall-clock time by 20.9\% relative to disabling both modules. This supports the role of local-state collection in making repairs more efficient.

\smallskip
\noindent{\bf Multi-error collection}. Removing multi-error collection 
mainly helps on cost savings. 
Full \tool has 7.0\% lower monetary cost and 10.9\% lower wall-clock time than the no-multi-error variant. When local-state collection is absent, keeping multi-error collection still reduces monetary cost by 11.6\% and wall-clock time by 2.8\% relative to disabling both modules.

\smallskip
\noindent{\bf Combined effect}. Disabling both modules gives the clearest degradation, reducing pass@10 to 92.0\% and trial-level success to 86.8\%. On the efficiency aspect, full \tool has 25.6\% lower monetary cost and 30.9\% lower wall-clock time than the variant with both modules disabled.

\begin{tcolorbox}[
  colback=gray!5,
  colframe=black!60,
  title=Response to RQ3
]
Both proof-specific feedback modules contribute to \tool's overall performance. Local-state collection has a clear effect on the proof success rate, while multi-error collection mainly reduces wasted repair cost. 
\end{tcolorbox}


\subsection{RQ4: Generalization to a Recent Project}
Finally, we evaluate \tool on the recent Parcas benchmark. We run each system 10 times using DeepSeek, 
and Table~\ref{tab:rq4-parcas} shows the results.

\tool keeps its advantage on the recent Parcas project. \tool solves 98/100 theorems at pass@10 with a 93.3\% trial-level success rate, compared with 94/100 and 88.4\% for Claude Code. OpenCode matches \tool's pass@10 at 98/100, but has lower trial-level success, 90.1\%. On the cost of shared-solved theorems, \tool reduces monetary cost by 52.8\% and wall-clock time by 25.5\% relative to Claude Code, and reduces monetary cost by 32.0\% and wall-clock time by 6.8\% relative to OpenCode.

\begin{tcolorbox}[
  colback=gray!5,
  colframe=black!60,
  title=Response to RQ4
]
\tool remains effective (over 90\% prove success rate) on the recent Parcas repository. It further reduces monetary cost and wall-clock time compared with Claude Code and OpenCode.
\end{tcolorbox}


\begin{table}[t]
\centering
\footnotesize
\caption{Ablation results of local-state collection and multi-error collection.} 
\label{tab:ablation-main}
\textbf{(a) Success rate comparison}\par\vspace{0.15em}
\begin{tabular}{@{}lcc@{}}
\toprule
Variant & Pass@10 & Trial Success \\
\midrule
Full RAIN & 95.0\% & 90.9\% \\
w/o local-state & 94.0\% & 88.0\% \\
w/o multi-error & 95.0\% & 91.4\% \\
w/o both & 92.0\% & 86.8\% \\
\bottomrule
\end{tabular}

\vspace{1em}
\textbf{(b) Cost comparison}\par\vspace{0.15em}
\begin{tabular}{@{}lr|rr@{}}
\toprule
Pair & Thm. & {Cost Red.} & {Time Red.} \\
\midrule
Full vs. w/o local-state & 94 & 13.0\% & 26.8\% \\
Full vs. w/o multi-error & 95  & 7.0\% & 10.9\% \\
Full vs. w/o both & 92 &  25.6\% & 30.9\% \\
w/o local-state vs. w/o both & 92 & 11.6\% & 2.8\% \\
w/o multi-error vs. w/o both & 92 & 21.4\% & 20.9\% \\

\bottomrule
\end{tabular}
\end{table}
\begin{table}[t]
\centering
\footnotesize
\caption{Parcas results using DeepSeek.}
\label{tab:rq4-parcas}
\textbf{(a) Success rate comparison}\par\vspace{0.15em}
\begin{tabular}{@{}lcc@{}}
\toprule
System & Pass@10 & Trial Success \\
\midrule
RAIN & 98/100 (98.0\%) & 93.3\% \\
Claude Code & 94/100 (94.0\%) & 88.4\% \\
OpenCode & 98/100 (98.0\%) & 90.1\% \\
\bottomrule
\end{tabular}\par
\vspace{1em}
\textbf{(b) Cost comparison}\par\vspace{0.15em}
\begin{tabular}{@{}lr|rr@{}}
\toprule
Pair & Thm. & {Cost Red.} & {Time Red.} \\
\midrule
RAIN vs. Claude Code & 94 & 52.8\% & 25.5\% \\
RAIN vs. OpenCode & 98 & 32.0\% & 6.8\% \\
\bottomrule
\end{tabular}
\end{table}

\subsection{Result Analysis}

\begin{figure}[t]
\centering
\includegraphics[width=\columnwidth]{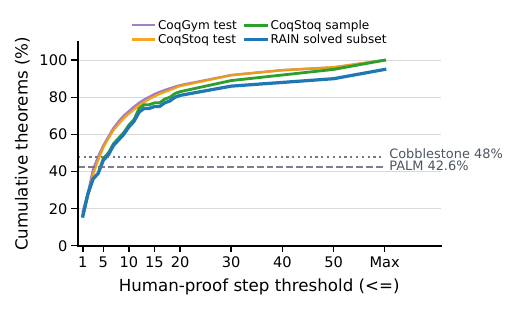}
\caption{Cumulative human-proof step distributions. 
PALM and Cobblestone lines are from different test sets, and this figure is merely intended to show the prove success rate trend against the proof length.} 
\label{fig:proof-step-cdf}
\Description{Cumulative distributions of human-proof lengths for the benchmark test sets, the sampled CoqStoq targets, and RAIN's solved subset. Most proofs contain at most thirty steps, with a small tail of longer proofs.}
\end{figure}

\noindent{\bf Proof-Step Distribution}.
This analysis relates the aggregate success rates to the human-proof lengths of the target theorems. Cobblestone and PALM both report results on CoqGym-derived settings, and Figure~\ref{fig:proof-step-cdf} compares CoqGym's test split with the CoqStoq test split used in our main experiments and with \tool's solved subset. The benchmark distributions are similar and heavily skewed toward short human proofs: about 92\% of theorems at thirty steps or below and about 96\% at fifty steps or below. Our 100-target sample is less concentrated in very short proofs, but still is dominated by short and medium proofs, with 89 targets requiring at most 30 steps and 95 requiring at most 50 steps.

\smallskip
\noindent{\bf Long-Proof Tail}.
The distribution matters for interpreting aggregate success rates. On uniformly sampled subsets from benchmarks with this profile, an improvement over Rango's reported 32\% on CoqStoq \cite{thompson2024rango} to around 70\% can still leave much of the long-proof region unresolved. The strict 50-step tail is small in our sample: according to CoqStoq's step counts, only 5 targets are above fifty steps, and \tool solves all 5 under pass@10. To avoid drawing too much from only five targets, we also inspect the 11 targets above thirty steps. \tool solves 9 of these 11 under pass@10, but its trial-level success rate drops from 95.7\% on targets at thirty steps or below to 51.8\% above thirty steps. Long-horizon proof construction therefore remains a distinct failure mode.

\smallskip
\noindent{\bf Remaining Failures}.
In our 100-theorem sample, \tool solves 95 instances under pass@10; the five remaining failures all come from the fourcolor development. We believe this concentration reflects the compact Mathematical Components and ssreflect style used in fourcolor~\cite{gonthier2010introduction}, where proofs often require fluency with boolean reflection, ssreflect views, canonical structures and coercions, and specialized graph-coloring idioms. This suggests that these features remain weak points for the backend model.

\section{Discussion}


\noindent{\bf Benchmark exposure}. LLMs may have encountered established theorem-proving benchmarks or their underlying repositories during training. We mitigate this threat by reporting the RQ4 experiment on Parcas~\cite{moine2026parcas}, whose repository was first uploaded on May 29, 2026, after the April 24, 2026 release of the DeepSeek model used in that experiment. This timing reduces the likelihood of direct target-project exposure, although general proof-pattern exposure remains possible.

\smallskip
\noindent{\bf LLM stochasticity}. Agent performance can vary across runs even under the same environment and prompt. To reduce the effect of 
randomness, our main agent comparisons run each theorem 10 times and report both theorem-level pass@10 and trial-level success. The artifact also includes the code and run configurations needed to reproduce the evaluated settings.

\smallskip
\noindent\textbf{Lexical search and semantic retrieval.}
%
The LLM can actively inspect repository proofs and decide what to explore next, but the underlying repository tools still rely primarily on names, tokens, and source text. As a result, the LLM often has to guess a useful identifier before it can locate the appropriate lemma. ReCent-Prover points to a promising direction by retrieving lemmas from LLM-generated proof plans, natural-language descriptions, and semantic embeddings~\cite{sun2026reasoning}. Such semantic retrieval could be integrated with agent-directed repository exploration, allowing the LLM to move seamlessly between high-level semantic search and concrete source inspection before attempting the next proof step.
\section{Related Work}

\subsection{Verified Software Systems}

Interactive theorem proving has already been used to verify substantial software artifacts. CompCert~\cite{leroy2009formal} verifies the semantic preservation of a realistic optimizing C compiler, while seL4~\cite{klein2009sel4} gives a machine-checked functional-correctness proof for an operating-system microkernel. Other systems-level verification projects cover different parts of the software stack: CertiKOS~\cite{gu2016certikos} verifies a layered concurrent OS kernel, FSCQ~\cite{chen2015using} verifies a crash-safe file system using Crash Hoare Logic, and seKVM~\cite{li2021formally} verifies memory protection for a commodity multiprocessor hypervisor. These systems show the value of ITPs for high-assurance software, but also expose the core bottleneck that motivates \tool: maintaining such proofs still requires substantial proof-engineering expertise, human effort, and time.

\subsection{Automated Theorem Proving}

Classical automated theorem proving provides much of the automation inside modern ITPs. SMT solvers such as Z3~\cite{de2008z3} and CVC4~\cite{barrett2011cvc4}, and first-order provers such as Vampire~\cite{riazanov2002design} and E~\cite{schulz2019faster}, are widely used backend engines for discharging proof goals. Hammer systems connect these engines to interactive theorem provers: Sledgehammer~\cite{blanchette2013extending} for Isabelle/HOL and CoqHammer~\cite{czajka2018hammer} for Rocq select relevant premises, translate goals to backend provers, and reconstruct accepted proof scripts when a backend succeeds. Hammers are strong local solvers, but they are not complete proof engineers; induction, case analysis, repository-specific tactics, and long proof plans often remain outside their direct reach.

Learning-based and LLM-based provers try to close this gap. Earlier systems such as ProverBot9001~\cite{sanchez2020generating} and Tactician~\cite{blaauwbroek2020tactician} learn proof-state-to-tactic behavior from proof corpora. Recent proof-generation systems combine generated scripts with repair, retrieval, decomposition, or checker feedback. In Rocq, PALM~\cite{lu2024proof} repairs LLM-generated scripts with error-specific mechanisms and backtracking, Rango~\cite{thompson2024rango} retrieves project-local proofs and lemmas at each proof step, Cobblestone~\cite{kasibatla2024cobblestone} recursively reuses successful parts of whole-proof attempts, and ReCent-Prover~\cite{sun2026reasoning} adds planning-based retrieval and reflection-based validation. Baldur~\cite{first2023baldur} applies whole-proof generation and error-guided repair in Isabelle/HOL. Other work targets complementary proof-search components: Thor~\cite{jiang2022thor} learns when to invoke Sledgehammer, Draft--Sketch--Prove~\cite{jiang2022draft} turns informal proof guidance into formal sketches whose holes are closed by hammers, LEGO-Prover~\cite{wang2024lego} grows helper-lemma libraries, Magnushammer~\cite{mikula2024magnushammer} learns transformer-based premise selection, LeanDojo/ReProver~\cite{yang2023leandojo} uses learned retrieval over accessible Lean premises, and BFS-Prover~\cite{xin2025bfs} scales best-first tactic search. Systems-verification studies such as LLM4FSCQ~\cite{qin2025can} and the seL4 neuro-symbolic proof-search work~\cite{he2026stepwise} further indicate that project-specific tactics, premises, and long procedural proof scripts remain major obstacles for off-the-shelf LLMs. Across these systems, several design choices recur: fixed premise retrieval, rule-based repair, hammer-backed local solving, and tactic-level proof-state search. \tool uses these techniques as motivation but treats proof generation as a repository-level agent workflow in which the LLM iteratively explores the project, edits the same proof script, and revises it using enriched checker feedback.

\subsection{Theorem-Proving Agents}

Recent Lean systems increasingly expose theorem proving through agentic interfaces. OpenGauss~\cite{mathinc2026opengauss} provides project-scoped Lean workflows, while Numina-Lean-Agent~\cite{liu2026numina} connects Claude Code to Lean through MCP-based proof tools. These systems show that modern code-agent workflows can be adapted to formal proof development. \tool follows the same broad direction for Rocq: it gives the LLM repository-level read and write tools, and further enhances checker feedback for proof repair.

\subsection{End-to-End Trusted Library Generation}

Proof agents can also be components of larger trusted-software pipelines. Fang and Xiong~\cite{fang2026trustworthy} study end-to-end trusted software project generation in Rocq. This direction broadens the goal from proving inside an existing repository to producing a trusted software artifact end to end. \tool focuses only on proof generation within this pipeline. 

\section{Conclusion}

This paper studied Rocq proof generation from the perspective of modern coding agents. Existing proof-automation systems have made important progress through retrieval, proof repair, hammers, decomposition, and tactic-level search. Modern coding agents provide a complementary workflow: they can iteratively inspect a project, edit a proof, run the checker, and revise from feedback. This workflow is well suited to formal proving because it resembles ordinary proof engineering while preserving the checker as the authority.

\tool shows that proof-specific agent design can make this workflow more effective and economical. 
The evaluation shows that \tool improves over prior Rocq baselines, stays competitive with strong coding agents, reduces cost, and remains effective on a recent Rocq project.


Future work may extend RAIN in two directions. First, combine RAIN with proof-decomposition methods for longer proofs, using RAIN as a leaf-goal prover for generated subgoals and helper lemmas. Second, use hybrid backends in RAIN, with cheaper specialized models handling routine retrieval and tactic selection while stronger LLMs handling difficult proof repair rounds.

\bibliographystyle{ACM-Reference-Format}
\bibliography{references}

@misc{milmo2024windows,
  author = {{The Guardian}},
  title = {{Windows} Global {IT} Outage: What We Know So Far},
  howpublished = {News article},
  month = jul,
  year = {2024},
  url = {https://www.theguardian.com/technology/article/2024/jul/19/windows-global-it-outage-what-we-know-so-far}
}

@misc{sabin2025cloudflare,
  author = {Sabin, Sam},
  title = {{Cloudflare} Outage Traced to One Oversized File},
  howpublished = {Axios},
  month = nov,
  year = {2025},
  url = {https://www.axios.com/2025/11/18/cloudflare-outage-cause-systems-down}
}

@article{gonthier2008formal,
  title={Formal proof--the four-color theorem},
  author={Gonthier, Georges and others},
  journal={Notices of the AMS},
  volume={55},
  number={11},
  pages={1382--1393},
  year={2008}
}

@article{leroy2009formal,
  title={Formal verification of a realistic compiler},
  author={Leroy, Xavier},
  journal={Communications of the ACM},
  volume={52},
  number={7},
  pages={107--115},
  year={2009},
  publisher={ACM New York, NY, USA}
}

@inproceedings{klein2009sel4,
  title={seL4: Formal verification of an OS kernel},
  author={Klein, Gerwin and Elphinstone, Kevin and Heiser, Gernot and Andronick, June and Cock, David and Derrin, Philip and Elkaduwe, Dhammika and Engelhardt, Kai and Kolanski, Rafal and Norrish, Michael and others},
  booktitle={Proceedings of the ACM SIGOPS 22nd symposium on Operating systems principles},
  pages={207--220},
  year={2009}
}

@inproceedings{sanchez2020generating,
  title={Generating correctness proofs with neural networks},
  author={Sanchez-Stern, Alex and Alhessi, Yousef and Saul, Lawrence and Lerner, Sorin},
  booktitle={Proceedings of the 4th ACM SIGPLAN International Workshop on Machine Learning and Programming Languages},
  pages={1--10},
  year={2020}
}

@inproceedings{lu2024proof,
  title={Proof automation with large language models},
  author={Lu, Minghai and Delaware, Benjamin and Zhang, Tianyi},
  booktitle={Proceedings of the 39th IEEE/ACM International Conference on Automated Software Engineering},
  pages={1509--1520},
  year={2024}
}

@article{czajka2018hammer,
  title={Hammer for Coq: Automation for dependent type theory},
  author={Czajka, {\L}ukasz and Kaliszyk, Cezary},
  journal={Journal of automated reasoning},
  volume={61},
  number={1},
  pages={423--453},
  year={2018},
  publisher={Springer}
}

@article{blanchette2013extending,
  title={Extending Sledgehammer with SMT solvers},
  author={Blanchette, Jasmin Christian and B{\"o}hme, Sascha and Paulson, Lawrence C},
  journal={Journal of automated reasoning},
  volume={51},
  number={1},
  pages={109--128},
  year={2013},
  publisher={Springer}
}

@inproceedings{blaauwbroek2020tactician,
  title={The tactician: A seamless, interactive tactic learner and prover for coq},
  author={Blaauwbroek, Lasse and Urban, Josef and Geuvers, Herman},
  booktitle={International Conference on Intelligent Computer Mathematics},
  pages={271--277},
  year={2020},
  organization={Springer}
}

@inproceedings{first2023baldur,
  title={Baldur: Whole-proof generation and repair with large language models},
  author={First, Emily and Rabe, Markus N and Ringer, Talia and Brun, Yuriy},
  booktitle={Proceedings of the 31st ACM Joint European Software Engineering Conference and Symposium on the Foundations of Software Engineering},
  pages={1229--1241},
  year={2023}
}

@article{thompson2024rango,
  title={Rango: Adaptive retrieval-augmented proving for automated software verification},
  author={Thompson, Kyle and Saavedra, Nuno and Carrott, Pedro and Fisher, Kevin and Sanchez-Stern, Alex and Brun, Yuriy and Ferreira, Jo{\~a}o F and Lerner, Sorin and First, Emily},
  journal={arXiv preprint arXiv:2412.14063},
  year={2024}
}

@article{kasibatla2024cobblestone,
  title={Cobblestone: A Divide-and-Conquer Approach for Automating Formal Verification},
  author={Kasibatla, Saketh Ram and Agarwal, Arpan and Brun, Yuriy and Lerner, Sorin and Ringer, Talia and First, Emily},
  journal={arXiv preprint arXiv:2410.19940},
  year={2024}
}

@article{thakur2023context,
  title={An in-context learning agent for formal theorem-proving},
  author={Thakur, Amitayush and Tsoukalas, George and Wen, Yeming and Xin, Jimmy and Chaudhuri, Swarat},
  journal={arXiv preprint arXiv:2310.04353},
  year={2023}
}

@article{yang2023leandojo,
  title={Leandojo: Theorem proving with retrieval-augmented language models},
  author={Yang, Kaiyu and Swope, Aidan and Gu, Alex and Chalamala, Rahul and Song, Peiyang and Yu, Shixing and Godil, Saad and Prenger, Ryan J and Anandkumar, Animashree},
  journal={Advances in Neural Information Processing Systems},
  volume={36},
  pages={21573--21612},
  year={2023}
}

@misc{rocqrefman900,
  author = {{The Rocq Development Team}},
  title = {The {Rocq} Prover Reference Manual, Version 9.0.0},
  howpublished = {Reference manual},
  year = {2025},
  url = {https://rocq-prover.org/doc/V9.0.0/refman/}
}

@article{sun2026reasoning,
  title={On Reasoning-Centric LLM-based Automated Theorem Proving},
  author={Sun, Yican and Shi, Chengwei and Lyu, Hangzhou and Xiong, Yingfei},
  journal={arXiv preprint arXiv:2604.19558},
  year={2026}
}

@inproceedings{xin2025bfs,
  title={Bfs-prover: Scalable best-first tree search for llm-based automatic theorem proving},
  author={Xin, Ran and Xi, Chenguang and Yang, Jie and Chen, Feng and Wu, Hang and Xiao, Xia and Sun, Yifan and Zheng, Shen and Ding, Ming},
  booktitle={Proceedings of the 63rd Annual Meeting of the Association for Computational Linguistics (Volume 1: Long Papers)},
  pages={32588--32599},
  year={2025}
}

@article{he2026stepwise,
  title={Stepwise: Neuro-Symbolic Proof Search for Automated Systems Verification},
  author={He, Baoding and Li, Zenan and Sun, Wei and Yao, Yuan and Chen, Taolue and Ma, Xiaoxing and Su, Zhendong},
  journal={arXiv e-prints},
  pages={arXiv--2603},
  year={2026}
}

@misc{vsrocq_placeholder,
  author = {{The VsRocq Development Team}},
  title = {{VsRocq}: {Visual Studio Code} Support for {Coq} and {Rocq}},
  howpublished = {Project repository},
  year = {2025},
  url = {https://github.com/rocq-prover/vsrocq}
}

@misc{opencode,
  author = {{OpenCode}},
  title = {Agents},
  howpublished = {Project documentation},
  year = {2026},
  url = {https://opencode.ai/docs/agents}
}

@misc{anthropic2026claudecode,
  author = {{Anthropic}},
  title = {{Claude Code} Overview},
  howpublished = {Product documentation},
  year = {2026},
  url = {https://code.claude.com/docs/en/overview}
}

@misc{openai2026codex,
  author = {{OpenAI}},
  title = {{Codex}},
  howpublished = {Developer documentation},
  year = {2026},
  url = {https://developers.openai.com/codex}
}

@misc{openai2026models,
  author = {{OpenAI}},
  title = {Models},
  howpublished = {{API} documentation},
  year = {2026},
  url = {https://developers.openai.com/api/docs/models}
}

@misc{deepseek2026pricing,
  author = {{DeepSeek}},
  title = {Models and Pricing},
  howpublished = {{API} documentation},
  year = {2026},
  url = {https://api-docs.deepseek.com/quick_start/pricing}
}

@misc{qwen3535ba3b,
  author = {{Qwen Team}},
  title = {{Qwen3.5-35B-A3B} Model Card},
  year = {2026},
  url = {https://huggingface.co/Qwen/Qwen3.5-35B-A3B}
}

@misc{moine2026parcas,
  author = {Moine, Antoine},
  title = {{Parcas}: A Separation Logic for Parallel Time Complexity with Work and Span Credits},
  howpublished = {Project repository},
  year = {2026},
  url = {https://github.com/nobrakal/parcas}
}

@article{gonthier2010introduction,
  title={An introduction to small scale reflection in Coq},
  author={Gonthier, Georges and Mahboubi, Assia},
  journal={Journal of formalized reasoning},
  volume={3},
  number={2},
  pages={95--152},
  year={2010}
}

@inproceedings{gu2016certikos,
  title={$\{$CertiKOS$\}$: An extensible architecture for building certified concurrent $\{$OS$\}$ kernels},
  author={Gu, Ronghui and Shao, Zhong and Chen, Hao and Wu, Xiongnan Newman and Kim, Jieung and Sj{\"o}berg, Vilhelm and Costanzo, David},
  booktitle={12th USENIX Symposium on Operating Systems Design and Implementation (OSDI 16)},
  pages={653--669},
  year={2016}
}

@inproceedings{chen2015using,
  title={Using Crash Hoare logic for certifying the FSCQ file system},
  author={Chen, Haogang and Ziegler, Daniel and Chajed, Tej and Chlipala, Adam and Kaashoek, M Frans and Zeldovich, Nickolai},
  booktitle={Proceedings of the 25th Symposium on Operating Systems Principles},
  pages={18--37},
  year={2015}
}

@inproceedings{li2021formally,
  title={Formally verified memory protection for a commodity multiprocessor hypervisor},
  author={Li, Shih-Wei and Li, Xupeng and Gu, Ronghui and Nieh, Jason and Hui, John Zhuang},
  booktitle={30th USENIX Security Symposium (USENIX Security 21)},
  pages={3953--3970},
  year={2021}
}

@inproceedings{de2008z3,
  title={Z3: An efficient SMT solver},
  author={De Moura, Leonardo and Bj{\o}rner, Nikolaj},
  booktitle={International conference on Tools and Algorithms for the Construction and Analysis of Systems},
  pages={337--340},
  year={2008},
  organization={Springer}
}

@inproceedings{barrett2011cvc4,
  title={cvc4},
  author={Barrett, Clark and Conway, Christopher L and Deters, Morgan and Hadarean, Liana and Jovanovi{\'c}, Dejan and King, Tim and Reynolds, Andrew and Tinelli, Cesare},
  booktitle={International Conference on Computer Aided Verification},
  pages={171--177},
  year={2011},
  organization={Springer}
}

@article{riazanov2002design,
  title={The design and implementation of VAMPIRE},
  author={Riazanov, Alexandre and Voronkov, Andrei},
  journal={AI communications},
  volume={15},
  number={2-3},
  pages={91--110},
  year={2002},
  publisher={SAGE Publications Sage UK: London, England}
}

@inproceedings{schulz2019faster,
  title={Faster, higher, stronger: E 2.3},
  author={Schulz, Stephan and Cruanes, Simon and Vukmirovi{\'c}, Petar},
  booktitle={International Conference on Automated Deduction},
  pages={495--507},
  year={2019},
  organization={Springer}
}

@inproceedings{qin2025can,
  title={Can Large Language Models Verify System Software? A Case Study Using FSCQ as a Benchmark},
  author={Qin, Jianxing and Du, Alexander and Zhang, Danfeng and Lentz, Matthew and Zhuo, Danyang},
  booktitle={Proceedings of the 2025 Workshop on Hot Topics in Operating Systems},
  pages={34--41},
  year={2025}
}

@article{jiang2022thor,
  title={Thor: Wielding hammers to integrate language models and automated theorem provers},
  author={Jiang, Albert Qiaochu and Li, Wenda and Tworkowski, Szymon and Czechowski, Konrad and Odrzyg{\'o}{\'z}d{\'z}, Tomasz and Mi{\l}o{\'s}, Piotr and Wu, Yuhuai and Jamnik, Mateja},
  journal={Advances in Neural Information Processing Systems},
  volume={35},
  pages={8360--8373},
  year={2022}
}

@article{jiang2022draft,
  title={Draft, sketch, and prove: Guiding formal theorem provers with informal proofs},
  author={Jiang, Albert Q and Welleck, Sean and Zhou, Jin Peng and Li, Wenda and Liu, Jiacheng and Jamnik, Mateja and Lacroix, Timoth{\'e}e and Wu, Yuhuai and Lample, Guillaume},
  journal={arXiv preprint arXiv:2210.12283},
  year={2022}
}

@inproceedings{wang2024lego,
  title={Lego-prover: Neural theorem proving with growing libraries},
  author={Wang, Haiming and Xin, Huajian and Zheng, Chuanyang and Liu, Zhengying and Cao, Qingxing and Huang, Yinya and Xiong, Jing and Shi, Han and Xie, Enze and Yin, Jian and others},
  booktitle={International Conference on Learning Representations},
  volume={2024},
  pages={31566--31597},
  year={2024}
}

@inproceedings{mikula2024magnushammer,
  title={Magnushammer: A transformer-based approach to premise selection},
  author={Miku{\l}a, Maciej and Tworkowski, Szymon and Antoniak, Szymon and Piotrowski, Bartosz and Jiang, Qiaochu and Zhou, Jin and Szegedy, Christian and Kuci{\'n}ski, {\L}ukasz and Mi{\l}o{\'s}, Piotr and Wu, Yuhuai},
  booktitle={International Conference on Learning Representations},
  volume={2024},
  pages={39326--39350},
  year={2024}
}

@misc{mathinc2026opengauss,
  author = {{Math, Inc.}},
  title = {{OpenGauss}: A Multi-Agent Workflow Orchestrator for {Lean}},
  howpublished = {Project website},
  year = {2026},
  url = {https://www.math.inc/opengauss}
}

@article{liu2026numina,
  title={Numina-Lean-Agent: An Open and General Agentic Reasoning System for Formal Mathematics},
  author={Liu, Junqi and Zhou, Zihao and Zhu, Zekai and Santos, Marco Dos and He, Weikun and Liu, Jiawei and Wang, Ran and Xie, Yunzhou and Zhao, Junqiao and Wang, Qiufeng and others},
  journal={arXiv preprint arXiv:2601.14027},
  year={2026}
}

@article{fang2026trustworthy,
  title={Trustworthy Software Project Generation: a Case Study with an Interactive Theorem Prover},
  author={Fang, Jian and Xiong, Yingfei},
  journal={arXiv preprint arXiv:2605.26017},
  year={2026}
}

@article{yao2022react,
  title={React: Synergizing reasoning and acting in language models},
  author={Yao, Shunyu and Zhao, Jeffrey and Yu, Dian and Du, Nan and Shafran, Izhak and Narasimhan, Karthik and Cao, Yuan},
  journal={arXiv preprint arXiv:2210.03629},
  year={2022}
}

\end{document}